\documentclass[twocolumn,showpacs,preprintnumbers,superscriptaddress,amsmath,amssymb,prl]{revtex4}

\usepackage{graphicx}
\usepackage{dcolumn}
\usepackage{bm}
\usepackage{amsmath}
\usepackage{indentfirst}

\newcommand{\be}{\begin{equation}}
\newcommand{\ee}{\end{equation}}
\newcommand{\ba}{\begin{eqnarray}}
\newcommand{\ea}{\end{eqnarray}}
\newcommand{\bs}{\boldsymbol}

\begin{document}

\preprint{APS preprint}

\title{Power Law Distributions of Seismic Rates}

\author{A. Saichev}
\affiliation{Mathematical Department,
Nizhny Novgorod State University, Gagarin prosp. 23,
Nizhny Novgorod, 603950, Russia}
\affiliation{Institute of Geophysics and Planetary Physics,
University of California, Los Angeles, CA 90095}

\author{D. Sornette}
\affiliation{Institute of Geophysics and Planetary Physics
and Department of Earth and Space Sciences,
University of California, Los Angeles, CA 90095}
\affiliation{Laboratoire de Physique de la Mati\`ere Condens\'ee,
CNRS UMR 6622 and Universit\'e de Nice-Sophia Antipolis, 06108
Nice Cedex 2, France}
\email{sornette@moho.ess.ucla.edu}

\date{\today}

\begin{abstract}
We report an empirical determination of the
probability density functions $P_{\text{data}}(r)$
of the number $r$ of earthquakes in finite space-time windows for the 
California catalog. We find 
a stable power law tail $P_{\text{data}}(r) \sim 1/r^{1+\mu}$ with exponent 
$\mu \approx 1.6$ for all space ($5 \times 5$ to $20 \times 20$ km$^2$)
and time intervals ($0.1$ to $1000$ days). These observations, as well
as the non-universal dependence on space-time windows for all different 
space-time windows simultaneously, are explained by solving one 
of the most used reference model in seismology (ETAS), which
assumes that each earthquake can trigger other earthquakes. 
The data imposes that active seismic regions are 
Cauchy-like fractals, whose exponent $\delta =0.1 \pm 0.1$ is well-constrained
by the seismic rate data.
\end{abstract}

\pacs{91.30.Px ; 89.75.Da; 05.40.-a}

\maketitle

Seismicity is perhaps the best example of a self-organizing process
exhibiting scaling diagnosed with so many power laws: the
Gutenberg-Richter distribution $\sim 1/E^{1+\beta}$ (with $\beta \approx
2/3$) of earthquake energies $E$; the Omori law $\sim 1/t^p$ (with $p
\approx 1$ for large earthquakes) of the rate of aftershocks as a 
function of time $t$ since a mainshock; the productivity law $\sim E^{a}$ (with
$a \approx 2/3$) giving the number of earthquakes triggered by an event of
energy $E$ \cite{H}; the power law distribution  $\sim 1/L^2$ of fault lengths
$L$ \cite{Davy}; the fractal structure of fault networks \cite{davy2} and of the spatial
organization of earthquake epicenters \cite{KK}; the distribution $1/s^{2+\delta}$
(with $\delta \geq 0$) of seismic stress sources $s$ in earthquake focal
zones due to past earthquakes \cite{kagan94}.
Related universal laws for the distribution
of waiting times and seismic rates between 
earthquakes have recently been derived from the analyses of 
space-time windows \cite{BaketalOmo}.

Here, we report and explain theoretically an addition empirical power law:
the numbers $r$ of earthquakes in finite
space-time windows for the California SCEC catalog, over fixed spatial
boxes $5 \times 5$ km$^2$ to $20 \times 20$ km$^2$ 
and time intervals $dt =1,~10,~100$ and $1000$
days are distributed with a stable power law tail $P_{\text{data}}(r)
\sim 1/r^{1+\mu}$ with exponent $\mu \approx 1.6$ for all time
intervals. This result has important implications in constraining
the physics of earthquakes and in estimating the performance of
forecasting models of seismicity. For the former, we show that
this observation can be rationalized by a simple stochastic
branching model (ETAS) which uses the above mentioned Gutenberg-Richter, Omori, and
productivity laws applied to a fractal spatial geometry
of earthquake epicenters. The fundamental physical ingredient is that
each earthquake can trigger other earthquakes (``aftershocks'') and an
earthquake sequence results in this model from the cascade of
aftershocks triggered by each past earthquake. In addition, the
growing efforts in earthquake forecasts requires estimating
the performance of a forecasting model by
a likelihood function, which are currently based on Poisson probabilities
calculated using declustered catalogues. Our work shows
that spontaneous fluctuations of the number of triggered earthquakes in
space-time bins, due to the cascades of triggering 
processes, may lead to dramatic departures from the Poisson model used
as one of the building block of standard testing procedures.

\begin{figure}[h]
\includegraphics[width=8cm]{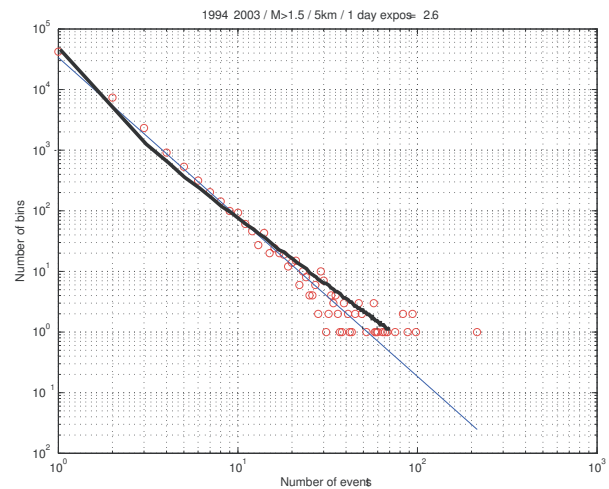}
\caption{\label{fig1} Empirical 
probability density functions (pdf) $P_{\text{data}}(r)$
of the number $r$ of earthquakes in the space-time bins of size $5 \times 5$
km$^2$ and $dt=1$ days of the SCEC catalog.
The straight line is the best fit with a pure power law (\ref{mfm,mfls}).
The continuous line is the theoretical prediction 
derived below for $dt=1$ days and with no
adjustable parameters, as the parameters are fixed to the values adjusted
from the fit for $dt=1000$ days for which the pdf is strongly curved.}
\end{figure}

In order to maximize the size and quality of the data used for our
analysis, we consider the time interval $1994-2003$ in a region
from approximately $32^\circ$ to $37^\circ$N in latitude, and from $-114^\circ$
to $-122^\circ$ in longitude, of the Southern Californian earthquakes
catalog with revised magnitudes $M_L>1.5$, which contains a total of $86,228$
earthquakes. The spatial domain is covered by square boxes of ($L=5$ km) $\times$ ($L=5$ km).
Other larger box sizes given similar results.
We consider time windows from $dt=1$ day to
$dt=1000$ days. Figure \ref{fig1} plots
the empirical probability density functions $P_{\text{data}}(r)$
of the number $r$ of earthquakes in the space-time bins described above
for  $dt=1$ days. The straight line is the best fit with a pure power law
\be
P_{\text{data}}(r) \sim 1/r^{1+\mu}
\label{mfm,mfls}
\ee
over the range $1 \leq r \leq 100$. Similar power law tails are present 
in the tail for the other time windows. 
The estimation for $\mu$ is stable since the fitted values are
$\mu = 1.65$ for $dt=100$ days, $\mu = 1.75$ for $dt=10$ days and
$\mu = 1.60$ for $dt=1$ days. However, the pdf
becomes more and more curved in the larger portion of the bulk as the size $dt$
of the time window is increased \cite{largepaper}. 
This behavior can be explained by the theory described below.

The ETAS (Epidemic-Type
Aftershock Sequence) model of triggered seismicity is based on the three
first well-founded empirical laws mentioned above. Its 
appeal lies in its simplicity, its power of explanation of other
empirical observations (see for instance
\cite{Forexp} and references therein) and its wide use as a benchmark.
The ETAS model belongs to the general class of branching processes
with infinite variance of the number of progenies per mother,
with a long-time (power law) memory of the impact of a mother on her
first-generation daughters described by the empirical Omori law for aftershocks.
These two ingredients together with the mechanism of cascades of branching
have been shown to give rise to subdiffusion
and to non mean-field behavior in the
distribution of the total number of aftershocks per
mainshock \cite{SaichHelmSor}, 
in the distribution of the total number of generations before
extinctions and in the
distribution of the total duration of an 
aftershock sequence before extinction \cite{SaichSorl04}.

In the ETAS model, each earthquake is a potential progenitor or mother, 
characterized by its conditional average number $N_m \equiv \kappa \mu(m)$
of children (triggered events or aftershocks of first generation), where
$\mu(m) = 10^{\alpha (m-m_0)}$ is proportional to the average 
productivity of an
earthquake of magnitude $m \geq m_0$ \cite{H}, $\kappa$ is a constant factor
and $m_0$ is the minimum magnitude of earthquakes capable of triggering
other earthquakes. For a given earthquake of magnitude $m$ and therefore
of mark $\mu(m)$, the number $r$ of its daughters of first generation
are drawn at random according to the Poissonian statistics
$p_\mu(r)= \frac{N_m^r}{r!}\,e^{-N_m} =
\frac{(\kappa\mu)^r}{r!}\,e^{-\kappa\mu}$. 
The challenge of our present analysis is to understand how the
exponential Poisson distribution is renormalized into the power law (\ref{mfm,mfls})
by taking into account all earthquake triggering paths simultaneously over all possible
generations. The ETAS model is complemented by the normalized Gutenberg-Richter (GR)
density distribution of earthquake magnitudes
$p(m) = b ~\ln (10)~ 10^{-b (m-m_0)}, m \geq m_0$. This magnitude
distribution $p(m)$ is assumed to be independent of the
magnitude of the triggering earthquake, i.e., a large
earthquake can be triggered by a smaller one.
Combining the GR and the productivity laws shows that the
earthquake marks $\mu$ and therefore the conditional average
number $N_m$ of daughters of first generation are
distributed according to the normalized power law
\be
p_{\mu}(\mu) = {\gamma \over \mu^{1+\gamma}}~,
~~~1 \leq \mu < +\infty, ~~~~~\gamma = b/\alpha~.
\label{aera}
\ee
For earthquakes, $b \approx 1$ and $0.5 < \alpha < 1$
giving $1 < \gamma <2$.
This range $1 < \gamma <2$ implies that the
mathematical expectation of $\mu$ and therefore of $N_m$
(performed over all possible magnitudes) is finite but its variance
is infinite. Given $\gamma$, the coefficient $\kappa$ then controls the value of the
average number $n$ (or branching ratio) of children of first generation per mother:
$n = \langle N_m \rangle = \kappa \langle \mu \rangle = \kappa {\gamma \over \gamma -1}$,
where the average $\langle N_m \rangle$ is taken over all mothers' magnitudes
drawn from the GR law. Recall that the values $n<1$, $n=1$ and $n>1$
correspond respectively to the sub-critical, critical and super-critical
branching regimes.
The last ingredient of the ETAS model consists in the specification of the
space-time rate function $N_m ~\Phi(\bs{r}-\bs{r}_i,t-t_i)$ giving 
the average rate of first generation daughters
at time $t$ and position $\bs{r}$ created by a mother of
magnitude $m\geq m_0$ occurring at time $t_i$ and position $\bs{r}_i$. We use
the standard factorization $\Phi(\bs{x},t)=\Phi(t)\, \phi(\bs{x})$.
The time propagator $\Phi(t)$ has the Omori law form
$\Phi(t)= \frac{\theta c^\theta}{(c+t)^{1+\theta}}~H(t)$
where $H(t)$ is the Heaviside function, $0<\theta<1$, $c$ 
is a regularizing time scale that
ensures that the seismicity rate remains finite close to
the mainshock. The space propagator is 
$\phi(\bs{x})= \frac{\eta ~d^{\eta}}{2 \pi (x^2+d^2)^{(\eta+2)/2}}$.
The next ingredient of the ETAS model is to assume that plate tectonic
motion induces spontaneous mother earthquakes, which are not triggered by 
previous earthquakes, according to a Poissonian point process, such
that the average number of spontaneous mother earthquakes per unit time
and per unit surface is $\varrho$. In the 
ETAS branching model, each such spontaneous mother earthquake then triggers 
independently its own space-time aftershocks branching process. 
The last ingredient of our theory is to recognize that, 
at large scale, earthquakes are 
preferentially clustered near the plate boundaries while, at smaller scales,
earthquakes are found mostly along faults and close to nodes between
several faults. We thus extend slightly the ETAS model to allow for 
the heterogeneity of the spontaneous earthquake sources $\varrho$ reflecting
the influence of pre-existing fault structures, some
rheological heterogeneity and complex spatial stress distributions. 
For this, we use the distribution of the stress field
in heterogeneous media and due to earthquakes \cite{kagan94}
which is found close to a Cauchy distribution. The simplest
prescription is to assume that $\varrho$ is itself random and distributed 
according to $\frac{1}{\langle\varrho\rangle}\,
f\left(\frac{\varrho}{\langle\varrho\rangle}\right)$, 
where $\langle\varrho\rangle$ is then statistical average of the random
space-time Poissonian source intensity $\varrho$. In the numerical
applications, we shall use the form
$f_\delta(x)= \frac{\delta+1}{\delta} \left(1+
\frac{x}{\delta}\right)^{-1-\delta}\qquad (\delta>0)$.

Due to the
independence between each sequence triggered by each spontaneous event, 
the generating probability function (GPF) of the number of
events (including mother earthquakes and all their aftershocks of all generations),
falling into the space-time window $\{[t,t+\tau]\times
\mathcal{S}\}$ is equal to
\begin{equation}\label{1}
\Theta_{\text{sp}}(z,\tau,\mathcal{S})=e^{-\varrho\,
L(z,\tau,\mathcal{S})}
\end{equation}
where $L(z,\tau,\mathcal{S})=\int_0^\infty dt
\iint\limits_{-\infty}^{\quad\infty} d\bs{x}
[1-\Theta(z,t,\tau,\mathcal{S};\bs{x})]+ \int_0^\tau dt
\iint\limits_{-\infty}^{\quad\infty} d\bs{x}
[1-\Theta(z,t,\mathcal{S};\bs{x})][1-I_\mathcal{S}(\bs{x})]+ 
\int_0^\tau dt \iint\limits_{\mathcal{S}}
d \bs{x}[1-z\Theta(z,t,\mathcal{S};\bs{x})]$.
The first summand in $L(z,\tau,\mathcal{S})$ describes the contribution to the GPF
$\Theta_{\text{sp}}$ from aftershocks triggered
by mother earthquakes that occurred before the time window
(i.e. at instants $t'$ such that $t'<t$). The corresponding GPF
$\Theta(z,t-t',\tau,\mathcal{S};\bs{x})$ of the
number of aftershocks triggered inside the space-time window
$\{[t,t+\tau]\times \mathcal{S}\}$ by some mother event that occurred at time $t'=0$
satisfies the relation
\begin{equation}
\label{3bb}
\Theta(z,t,\tau,\mathcal{S};\bs{x})=
G[1-\Psi(z,t,\tau,\mathcal{S};\bs{x})]\,,
\end{equation}
where the auxiliary function
$\Psi(z,t,\tau,\mathcal{S};\bs{x})$, describing the
space-time dissemination of aftershocks triggering by
some mother event, is equal to
\begin{equation}  
\label{4}
\begin{array}{c}
\displaystyle \Psi(z,t,\tau,\mathcal{S};\bs{x})= \Phi(\bs{x},t) \otimes [1-
\Theta(z,t,\tau,\mathcal{S};\bs{x})] \\[2mm]
+ \Phi(\bs{x},t+\tau)\otimes [1-
\Theta(z,\tau,\mathcal{S};\bs{x})]+
\\[2mm]
(1-z)\Phi(\bs{x},t+\tau)\otimes I_\mathcal{S}(\bs{x})
\Theta(z,\tau,\mathcal{S};\bs{x})\,.
\end{array}
\end{equation}
The function $I_\mathcal{S}(\bs{x})$ in (\ref{4}) is
the indicator of the space window $\mathcal{S}$ and $G(z)$ in (\ref{3bb})
is the GPF of the number $R_1$ given by (\ref{aera}) of first generation
aftershocks triggered by some mother earthquake, given by
$G(z)=\gamma\kappa^\gamma(1-z)^\gamma\,
\Gamma(-\gamma,\kappa(1-z))$. 
The last two summands of $L(z,\tau,\mathcal{S})$ describe the
contribution of aftershocks triggered by earthquakes,
occurring inside the time window (i.e., $t'\in[t,t+\tau]$).
The second (resp. third) term corresponds to the subset
spatially outside (resp. inside) the domain $\mathcal{S}$. 
These last two terms depend on the GPF $\Theta(z,\tau,\mathcal{S};\bs{x})=
\Theta(z,t=0,\tau,\mathcal{S};\bs{x})$
of the numbers of aftershocks triggered till time $\tau$ inside
the space window $\mathcal{S}$ by some mother event
arising at the instant $t=0$ and at the point $\bs{x}$.
It follows from (\ref{3bb}) and (\ref{4}) that it satisfies
the relations
\begin{equation}\label{5}
\Theta(z,\tau,\mathcal{S};\bs{x})=
G[1-\Psi(z,\tau,\mathcal{S};\bs{x})]
\end{equation}
and
\begin{equation}\label{6}
\begin{array}{c}
\Psi(z,\tau,\mathcal{S};\bs{x})=\Phi(\bs{x},\tau)\otimes [1-
\Theta(z,\tau,\mathcal{S};\bs{x})]\\
 + (1-z)\Phi(\bs{x},\tau)\otimes I_\mathcal{S}(\bs{x})
\Theta(z,\tau,\mathcal{S};\bs{x})\,.
\end{array}
\end{equation}

In addition, we shall need the GPF
\be
\Theta(z,\mathcal{S};\bs{x})=
\Theta(z,\tau=\infty,\mathcal{S};\bs{x})
\ee
of the total numbers of aftershocks triggered by some mother
earthquake inside the area $\mathcal{S}$. As seen
from (\ref{5}) and (\ref{6}), it satisfies the relations
\begin{equation}\label{7}
\Theta(z,\mathcal{S};\bs{x})=
G[1-\Psi(z,\mathcal{S};\bs{x})]
\end{equation}
and $\Psi(z,\mathcal{S};\bs{x})= 1- \phi(\bs{x})\otimes
\Theta(z,\mathcal{S};\bs{x})+ (1-z) \phi(\bs{x})\otimes
I_\mathcal{S}(\bs{x}) \Theta(z,\mathcal{S};\bs{x})$.

Taking into account the distribution of the source intensities $\varrho$
amounts to averaging equation (\ref{1}) over $\varrho$ weighted with the statistics 
$\frac{1}{\langle\varrho\rangle}\,
f\left(\frac{\varrho}{\langle\varrho\rangle}\right)$. This gives
\begin{equation}
\label{15}
\Theta_{\text{sp}}(z,\tau;\mathcal{S})=\hat{f}[\langle\varrho\rangle\,
L(z,\tau,\mathcal{S})]\,,
\end{equation}
where $\hat{f}(u)$ is the Laplace transform of the pdf $f(x)$.

To go further, we make two approximations.
If the time duration $\tau$ of the
space-time window is sufficiently large, it can be shown that 
the statistical averages of the seismic rates become independent 
of $\tau$. It seems reasonable to conjecture that the 
GPF $\Theta(z,\tau,\mathcal{S};\bs{x})$ of the total number
of aftershocks triggered by some earthquake source inside
the space domain $\mathcal{S}$ until time $\tau$ 
coincides approximately with the saturated GPF
$\Theta(z,\mathcal{S};\bs{x})$ of the total number of aftershocks
triggered by some earthquake source inside the space domain
$\mathcal{S}$. Within this approximation of large time windows, 
the effect of aftershocks triggered by
earthquake sources occurring till the beginning $t$ of the time window is
negligible. 
Within this large time window approximation, one may ignore the first term 
in the contribution to (\ref{1}) and replace 
$\Theta(z,t,\mathcal{S};\bs{x})$ by
$\Theta(z,\mathcal{S};\bs{x})$ in the remaining terms. As a result, \
$L(z,\tau,\mathcal{S})$ in (\ref{1})
takes the following approximate form
$L(z,\tau,\mathcal{S})\simeq  \tau \iint\limits_{-\infty}^{\quad\infty}
[1-\Theta(z,\mathcal{S};\bs{x})][1-I_\mathcal{S}(\bs{x})]
d \bs{x}+ \tau \iint\limits_{\mathcal{S}}[1-z\Theta(z,\mathcal{S};\bs{x})] d \bs{x}$,
where $\Theta(z,\mathcal{S};\bs{x})$ is the solution of 
$\Theta=G\left[\Theta\otimes\phi-(1-z)
I_\mathcal{S}\Theta\otimes \phi\right]$.
To find a reasonable approximate expression for the sought GPF
$\Theta(z,\mathcal{S};\bs{x})$, notice that if the spatial extend $\ell$ 
of the window
is larger that the characteristic scale $d$ of the space kernel,
or if $n$ is close to $1$, then the characteristic spatial
scale associated with the GPF
$\Theta(z,\mathcal{S};\bs{x})$ becomes greater than
$d$. Therefore, without essential error,
one may replace $\Theta\otimes\phi$ by
$\Theta$. In addition, we take into account
the finiteness of the domain $\mathcal{S}$ by using the factorization
procedure: $I_\mathcal{S}(\bs{x})\Theta(z,\mathcal{S};\bs{x})\otimes
\phi(\bs{x})\simeq \Theta(z,\mathcal{S};\bs{x})~
p_\mathcal{S}(\bs{x})$,
where $p_\mathcal{S}(\bs{x})$ remains to be specified.
This amounts to replacing a convolution integral by an algebraic term.
This factorization approximation is a crucial step of our analysis and
is justified elsewhere \cite{largepaper}.
As a result of its use, the nonlinear integral equation for $\Theta(z,\mathcal{S};\bs{x})$
transforms into the functional equation
$\Theta=G[(1+(z-1) p_\mathcal{S}(\bs{x}))\Theta]$. This approximation
leads to $\langle R\rangle= \frac{n}{1-n}\, p_\mathcal{S}$,
where $\langle R\rangle$ is the average of the total number of events in the
space-time window. The effective parameter $p_\mathcal{S}$
can be determined from the consistency condition such that $\langle R\rangle$
be equal to the true $\langle R(\mathcal{S};\bs{x})\rangle$, which can
be calculated exactly. This gives
\begin{equation}
\label{34}
p_\mathcal{S}(\bs{x})= \frac{1-n}{n} \langle
R(\mathcal{S};\bs{x})\rangle\,, ~~
\tilde{p}_\mathcal{S}(\bs{q})=\tilde{I}_\mathcal{S}(\bs{q})
\tilde{\phi}(\bs{q}) \frac{1-n
}{1-n\tilde{\phi}(\bs{q})}\,.
\end{equation}
For $\ell\gg d$, the factor
$p_\mathcal{S}(\bs{x})$ approaches a rectangular function.
We can use this observation to help determine the statistics of the 
number of events in a finite space-time window, using the
approximation $p_\mathcal{S}(\bs{x})\simeq \text{const}=p$
for $\bs{x}\in \mathcal{S}$. We define the parameter $p$ 
as the space average of $p_\mathcal{S}(\bs{x})$ over the window's
area $\mathcal{S}$:
\begin{equation}
\label{35}
p\simeq \frac{1}{S} \iint\limits_{\mathcal{S}}
p_\mathcal{S}(\bs{x}) d \bs{x}\,.
\end{equation}
This approximation allows us to get
\be
\tau \iint\limits_{\mathcal{S}}[1-z\Theta(z,\mathcal{S};\bs{x})]
d \bs{x} \simeq \tau S [1-z\Theta(z;p)]~,
\ee
where $\Theta(z,p)$ is the solution of 
$\Theta(z;p)= G[(1+(z-1)p)\Theta(z;p)]$.

Complementarily, a study of 
$p_\mathcal{S}(\bs{x})$ shows that it is small outside the window space domain
$\mathcal{S}$. This implies that, outside 
$\mathcal{S}$, one may replace the functional equation on $\Theta$ by
$1-\Theta(z,\mathcal{S};\bs{x})\simeq \frac{n}{1-n} (1-z) p_\mathcal{S}(x)$.
Therefore, we get
$\tau \iint\limits_{-\infty}^{\quad\infty}
[1-\Theta(z,\mathcal{S};\bs{x})][1-I_\mathcal{S}(\bs{x})]
d \bs{x}\simeq q \tau S\, \frac{n}{1-n} (1-z)$,
where $q= \frac{1}{S} \iint\limits_{-\infty}^{\quad\infty}
p_\mathcal{S}(\bs{x})[1-I_\mathcal{S}(\bs{x})]d\bs{x}$.
Taking into account that
$\iint\limits_{-\infty}^{\quad\infty}
p_\mathcal{S}(\bs{x})d\bs{x}= S$, we obtain $q \simeq 1-p$.
Putting all these 
approximations together allows us to rewrite the expression of 
$L(z,\tau,\mathcal{S})$ in (\ref{1}) as
\begin{equation}
\label{37}
L(z,\tau,\mathcal{S})\simeq \tau S \left[ \frac{n}{1-n}
(1-p) (1-z)+1-z \Theta(z;p)\right].
\end{equation}
The factorization procedure obtains the characteristic
features of the space-time branching process in a finite space-time domain,
at the cost of an adjustable parameter $p$. 

Starting from the general expression (\ref{15}) of the GPF
$\Theta_{\text{sp}}(z,\tau;\mathcal{S})$ with the approximation (\ref{37})
for $L(z,\tau,\mathcal{S})$ and using the relationship between the probability
$P_{\text{sp}}(r;\rho,p)$ and its GPF in the form of its
integral representation, we obtain the following
expression valid in the limit of sufficiently large time windows
$P_{\text{sp}}(r;\rho,p)= \frac{1}{2\pi i}
\oint\limits_{\mathcal{C}} \hat{f}\left(\rho\, \left[
\frac{n}{1-n} (1-p) (1-z)+1-z \Theta(z;p)\right]\right)
\frac{dz}{z^{r+1}}$.
Introducing the new integration variable  
$y= (1+(z-1)p)\Theta(z;p)\quad \iff\quad z=Z(y)=
\frac{1}{p} \left(\frac{y}{G(y)}+p-1\right)$,
by construction of $y$, $\Theta(z;p)=G(y)$ which allows us to obtain the
following explicit expression
$P_{\text{sp}}(r;\rho,p)= \frac{1}{2\pi i} \times 
\oint\limits_{\mathcal{C}'} \hat{f}\left(\rho\, \left[
\frac{n}{1-n} (1-p) (1-Z(y))+1-Z(y) G(y)\right]\right)
\frac{dZ(y)}{dy} \frac{dy}{Z^{r+1}(y)}$.
This expression makes a precise
quantitative prediction for the dependence of the distribution $P_{\text{sp}}(r;\rho,p)$
of the number $r$ of earthquakes per space-time window as a function of
$r$, once the following model parameters are given: 
the branching ratio $n$, the exponent 
$\gamma$ of the distribution of productivities, the exponent $\delta$ 
of the distribution of spontaneous earthquake sources, 
the fraction $p_\mathcal{S}$
of direct (first generation) aftershocks which fall within 
the domain $\mathcal{S}$, and
the average number $\rho$ of spontaneous earthquake source per space-time
bin defined by $\rho=\langle\varrho\rangle\,\tau S$. The
theoretical curve in Fig.~\ref{fig1} is
obtained by a numerical integration of $P_{\text{sp}}(r;\rho,p)=$ for the set
of parameters $n=0.96$, $\gamma = 1.1$, $p_\mathcal{S}=0.25$, $\delta = 0.15$
and $\rho=0.0019~ dt$ with $dt$ in units of days (thus equal 
to $100$ for Fig.~\ref{fig1} ). These parameters give the best fit
for the large time window $dt=1000$ days. They have been kept fixed for 
the other time windows which exhibit very different
shapes in their bulk. The theory is thus able to account simultaneously for
all the considered time windows, with no adjustable parameters for the three
smallest time windows \cite{largepaper}.

We would also like to stress that, according to our theory,
the value of the exponent $\mu \approx 1.6$ used
in (\ref{mfm,mfls}) to fit the tails of the distributions of seismic rates is
describing a cross-over rather than a genuine asymptotic tail. Recall that
the distribution of the total number of aftershocks has two power law
regimes $\sim 1/r^{1+{1 \over \gamma}}$ for $r < r^* \simeq 1/(1-n)^{\gamma/(\gamma -1)}$
and $\sim 1/r^{1+ \gamma}$ for $r>r^*$ \cite{SaichHelmSor}. The existence
of this cross-over together with
the concave shape of the distribution at small and intermediate values of $r$
combine to create an effective power law with an apparent exponent $\mu \approx 1.6$
larger than the largest asymptotic exponent $\gamma$.
We have verified this to be the case in synthetically generated distributions
with genuine asymptotics exponent $\gamma=1.25$ for instance, which could be 
well fitted by $\mu \approx 1.6$ over several decades. We note also that
Pisarenko and Golubeva \cite{PisGol}, with a different approach applied to much
larger spatial box sizes in California, Japan and Pamir-Tien Shan, have 
reported an exponent $\mu<1$ which could be associated with 
the intermediate asymptotics characterized by the exponent $1/\gamma < 1$, found
in our previous analysis \cite{SaichSorl04}. By using data collapse 
with varying spatial box sizes on a California catalog, 
Corral finds that the distribution of seismic
rates exhibits a double power-law behavior with $\mu \approx 0$ for small rates
and $\mu \approx 1.2$ for large rates \cite{BaketalOmo}.  The first regime
might be associated with the non universal bulk part of the distribution
found in our analysis. The second regime is compatible with the 
prediction for the asymptotic exponent $\mu = \gamma$.
In conclusion, we have offered a simple explanation of the power law
distribution of seismic rates, which is derived from the other known
power laws and the physics of cascades of earthquake triggering.

This work was partially supported by NSF-EAR02-30429 and by
the Southern California Earthquake Center funded by NSF.

\end{document}